# A controlled study of virtual reality in first-year magnetostatics

C. D. Porter[1]

[1]*Department of Physics, The Ohio State University, 191 W. Woodruff Ave, Columbus, OH, 43210*

J. Brown[2], J. R. Smith[1], A. Simmons[1], M. Nieberding[1], A. Ayers[3], C. Orban[1]

[2]*College of Engineering, The Ohio State University, 2070 Neil Ave., Columbus, OH, 43210*
[3]*Advanced Computing Center for the Arts and Design, The Ohio State University, 1813 N. High St.., Columbus, OH, 43210*

Stereoscopic virtual reality (VR) has experienced a resurgence due to flagship products such as the Oculus Rift, HTC Vive and smartphone-based VR solutions like Google Cardboard. This is causing the question to resurface: how can stereoscopic VR be useful in instruction, if at all, and what are the pedagogical best practices for its use? To address this, and to continue our work in this sphere, we performed a study of 289 introductory physics students who were sorted into three different treatment types: stereoscopic virtual reality, WebGL simulation, and static 2D images, each designed to provide information about magnetic fields and forces. Students were assessed using preliminary items designed to focus on heavily-3D systems. We report on assessment reliability, and on student performance. Overall, we find that students who used VR did not significantly outperform students using other treatment types. There were significant differences between sexes, as other studies have noted. Dependence on students' self-reported 3D videogame play was observed, in keeping with previous studies, but this dependence was not restricted to the VR treatment.

## I. INTRODUCTION

Many topics in physics are inherently three-dimensional (3D), but are usually taught using two-dimensional media such as whiteboards and computer screens. Stereoscopic virtual reality (VR) is a 3D medium that might be suitable for teaching some content in physics and other STEM disciplines.

In recent years, VR technology has advanced rapidly, and the past few years have seen an explosion of VR consumer products [1]. Inexpensive solutions have emerged that leverage the computational power of smartphones to create a VR experience, as in a *Google Cardboard* setup [2]. This setup requires only a smartphone and an investment of a few dollars for cardboard or plastic headsets. This minimal investment means each student can have their own VR headset, which allows each student more exposure time. Although this is logistically promising, there are open questions about how to best utilize VR in STEM education. Specifically, research has consistently shown that technology by itself does not do much to enhance students' learning if it is not integrated in the classroom based on sound educational theories [3,4].

Many prior studies have described the application of VR interventions in various STEM classes, and with varying degrees of success. Students given VR interventions have reported being more engaged with the material, or having a better conceptual understanding than control students, but the advantage of VR over other media in achieving gains in specific learning outcomes is still unclear [5-14]. Because of the prohibitive cost of conventional VR headsets, many of these prior studies have limited sample sizes and in some cases VR treatment was not compared to a control group. But there have been large studies with careful controls. Madden et al. considers a VR intervention for an astronomy course on the topic of the seasons [14]. Another study by Smith et al., which is co-authored by several authors of this paper, considered a VR intervention on the topic of electrostatics [15]. Both works found only small differences in gains between the VR treatment group and others. One explanation for this for the electrostatics study is that the intervention in Smith et al. was comparatively passive and non-interactive. A metastudy by Hundhausen et al. in [16] found that how students use VR is more important than what visualizations are displayed, and that VR is most effective when it is constructively interactive, meaning that the student can effect change in the virtual environment, not merely view it. VR-based classroom projects have led to large gains over control groups in the context of mathematics[17]. The present study uses interventions that are slightly more interactive than those used in Smith et al., and we consider a different area of physics: magnetostatics.

Due to a lack of independently-validated assessments for magnetism (see Methods section) with a high fraction of 3D questions, we developed a suite of questions as a preliminary survey of 3D magnetostatics. The reliability of this survey is discussed below, along with student performance.

## II. METHODS

As students entered the testing area, they were randomly assigned to one of three treatment types (VR, WebGL, and static 2D images). The assessments were identical for all students, regardless of treatment type, except for a few questions posed in the VR treatment itself which were used to ensure students were engaging with the treatment. Scores on those questions asked within the VR simulation are not discussed here. The students' average overall performance in physics was fairly constant between treatment types, as determined by post-hoc analysis of students' final scores as a percentage of points in their physics course (VR: 81%, WebGL: 81%, Images: 83%). There was slight variation in the percentage of students reporting their sex as female in the three treatment types (VR: 23%, WebGL: 15%, Images: 23%). All treatment types involved visualization of magnetic fields due to moving point-like charged particles or due to long current-carrying wires, and also the visualization of magnetic forces and torques.

### A. Treatments

**Virtual reality:** VR visualizations were created as Android smartphone applications. The apps were written using Unity, a cross-platform game engine developed by Unity Technologies [18], and the Google VR SDK for Unity. Existing 3D posable hands were used for the "right-hand rule" visualizations [19]. The application used in this portion of the VR study displayed magnetic field vectors due to a moving point-like charged particle and also due to long current-carrying wires. The magnetic field was represented as an array of vectors as opposed to using the density of continuous field lines. The application was then built as an Android application package (APK file), and installed on two OSU-owned Nexus 5X smartphones. The app splits the phone's screen into two halves, one for each eye. Each phone is then placed in a cardboard or plastic viewer. The students can then view the magnetic systems in stereoscopic 3D. The app utilizes the smartphone's sensors so that when the students turn their heads, the system being displayed on the phone rotates, allowing students to see it from any orientation. Students were shown 5 instructional scenes and were told to "look around" and study the magnetic field vectors from many angles before moving on. Students were also asked a series of 7 questions within the VR simulation to ensure that students were engaging with the content. Students controlled the rate at which the visualizations progressed.

**WebGL:** The same visualizations used in VR were also exported for use in a web browser (WebGL format). Students could rotate the systems by clicking and dragging with a mouse, and could advance scenes using the space bar. The only other difference between VR and WebGL treatments was that the screen was not split and the user was not "immersed" in the WebGL treatment.

**Images:** Students in this treatment group were shown static 2D images of moving charges and current-carrying wires and the magnetic fields around them. The images were taken both from textbooks and from screenshots of the WebGL simulations. Since these students were shown only the type of visualization found in their textbooks and shown to them in class, this "treatment" serves as a control group.

### B. Assessment

Discussions with experienced instructors were used to determine what aspects of magnetic fields and forces were most fundamental to student progress in general, and specifically in OSUs introductory electricity and magnetism course. From the general themes that arose, the study team selected a subset that can be highly three-dimensional in nature, and are therefore most likely to be aided by stereoscopic 3D treatments. The themes selected were

1. **The direction of magnetic fields due to simple sources (using the right-hand rule for currents)**
2. **Superposition of magnetic fields from two or more sources**
3. **Magnetic forces (using the right-hand rule for forces)**
4. **Magnetic torque**

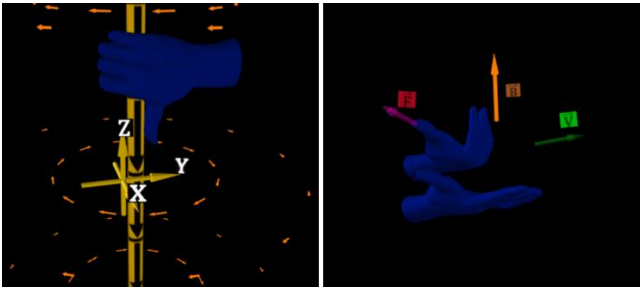

Fig. 1. Two scenes used in the VR and WebGL treatments. The left panel shows the right-hand rule for fields applied to a current-carrying wire. The right panel shows the right-hand rule for magnetic forces.

Treatment scenes were designed to illustrate critical aspects of the above themes, and to address common pitfalls. Two screenshots from example scenes are shown in Fig. 1.

Prior to any VR instruction, students were given a pretest that consisted of about 20 multiple choice questions on magnetic field directions, forces, and torques. Two example systems are shown in Fig. 2.

The post-test consisted of about 20 multiple choice questions. The multiple-choice questions were very similar to the pretest questions; in some cases they were verbatim repetitions, and in other cases they were modified slightly by asking for the field direction at a different point in space, or by reversing current direction.

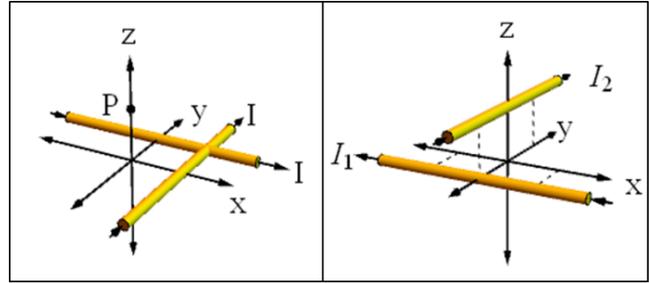

Fig. 2. System setups used in pretest and posttest questions. Students were asked for the direction of the magnetic field at point P (left), and for the direction of the net force and torque on the upper wire due to the lower wire (right).

From an experimental design perspective, it would have been preferable to assess learning using established validated metrics such as the Brief Electricity and Magnetism Survey (BEMA) [20] or the Conceptual Survey in Electricity and Magnetism (CSEM) [21]. However, review of existing validated assessments suggested that none are sufficiently focused on three-dimensional magnetostatics problems to be suited to this type of study. For example, the BEMA has only 4 questions (out of 30) that involve a system that is necessarily three-dimensional, or have a correct answer that involves a third dimension (into or out of the page). Since VR is most likely to be implemented and useful in areas that involve three dimensional systems, any practical assessment of student use of VR required a new assessment instrument focused on three-dimensional systems. The reliability of the preliminary assessment (in four subscales matching the four themes above) is discussed in the Results and Discussion section.

### C. Logistics

Participants were recruited from an introductory calculus-based electricity and magnetism course taken primarily by engineering and physical science majors. Students received points equivalent to one homework assignment for attending our testing session. Students were given informed consent forms, and those who declined to participate in a research study were given an alternative assignment and all their data were removed from our database. Approximately 96% of all students agreed to participate in research, resulting in 289 participants.

In addition to the pre-post questions on magnetism, students were asked the following background questions: "How often did you play video games when you were younger?", "How often do you play videogames now?" and "When you do play video games, are they primarily 2D or 3D?". In that last question, examples of 2D games such as Candy Crush were given, and examples of 3D games such as Minecraft were given. Students were also asked their sex and race.

## III. RESULTS & DISCUSSION

### A. Assessment reliability

The reliability of the subscales was assessed using Cronbach's alpha. The results are shown in Table I. Only one subscale showed high reliability (magnetic fields due to superpositions of two sources). All other subscales showed marginal or unacceptable reliability. The reliability measures were fairly consistent between pretest and posttest.

TABLE I. Cronbach's alpha for four subscales on the pre and posttests

| Theme | Alpha (pre) | Alpha (post) |
|---|---|---|
| 1 | 0.59 | 0.61 |
| 2 | 0.85 | 0.84 |
| 3 | 0.35 | 0.57 |
| 4 | 0.60 | 0.65 |

No subscale appeared to be at floor or ceiling in terms of student scores. The means and standard deviations are shown in Table II.

Table II. Mean percentages on the four subscales

| Theme | Pre Mean (%) | Post Mean (%) |
|---|---|---|
| 1 | 53±1 | 66±1 |
| 2 | 66±1 | 66±1 |
| 3 | 41±1 | 43±1 |
| 4 | 39±1 | 42±1 |

Future work should build upon the one well-behaved subscale. In the interests of both accurate reporting and adherence to a well-behaved subscale, in what follows, we report results on both the single well-behaved subscale and on the overall set of items.

### B. Assessment results by treatment

None of the treatment groups showed statistically significant gains from the pretest to posttest when all items are included in the score (see Fig. 3). This was tested using a repeated measures analysis in SPSS which yielded $p = 0.27$.

The lack of improvement is not entirely attributable to the poor reliability of the subscales, since there were also no statistically significant gains on the superposition subscale ($p > 0.5$). Adding the treatment type as a between-subjects factor in the repeated measures analysis again does not reveal statistically significant differences between treatment groups on all items ($p > 0.4$), nor on the superpositions subscale ($p > 0.8$).

The small to non-existent gains, and lack of dependence on treatment type make further analysis somewhat questionable. However, potential new treatments related to visuospatial rotations cannot completely ignore possible interactions with student sex and prior experience with visuospatial rotations in an electronic context. We therefore note that when both treatment type and student sex are added as between-subject factors, there is a statistically significant dependence on treatment, sex, and a statistically significant interaction between treatment and sex (with $p < 0.01$ in all cases). This is true both of the entire item set and of the superposition subscale. Although the sex question was asked using a write-in text box, no student reported a sex other than male or female. The results on all items are shown broken down by treatment and sex in Fig 4.

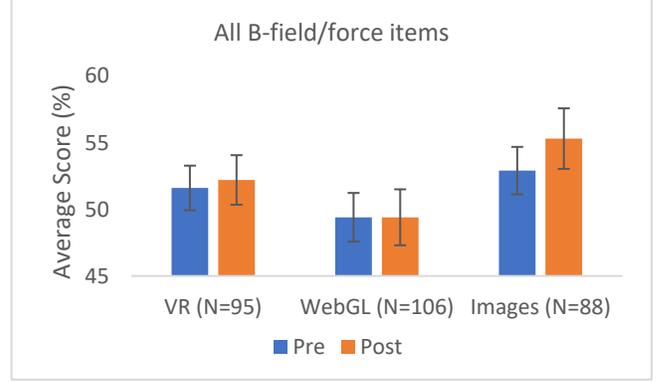

Fig. 3. Pre-post results for different treatment groups

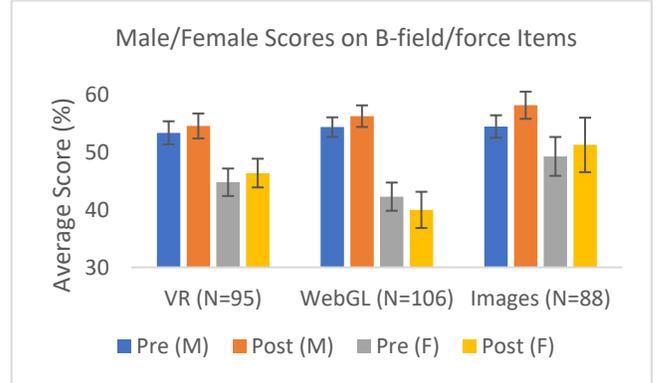

Fig. 4. Student scores on all items broken down by both treatment and sex.

One measure of interest is how different are male gains from female gains in the three treatments. If we call the gains for females in treatment $i$: $\left(g_{F,i}\right)_{av} = \left(\text{Post}_{F,i} - \text{Pre}_{F,i}\right)_{av}$, then the relevant effect size could be written as a comparison between gains by these two sexes in treatments $i$ and $j$:

$$d_{ij} = \frac{\left(g_{F,i} - g_{M,i}\right)_{av} - \left(g_{F,j} - g_{M,j}\right)_{av}}{SD}$$

The effect sizes are not trivial to calculate in this case, since each category being considered (sex, treatment) have different population sizes. In order to estimate the effect sizes, a simple bootstrapping (with replacement) technique was employed. Monte Carlo sampling of actual student scores were used to generate new data sets. These new data sets were used to determine the relevant standard deviation for use in the effect sizes. Convergence to two significant digits was reached after 100,000 bootstrapping samples were used. The resulting effect sizes are shown in Table III.

TABLE III. Effect sizes ($d$) for gender gap in gains by treatment

| Treatments | $d$ |
|---|---|
| VR-WebGL | 0.39 |
| VR-Images | 0.18 |
| WebGL-Images | 0.20 |

Here, the positive $d_{\text{VR-WebGL}}$, for example, indicates that women's gains beat men's gains by a greater amount in the VR treatment than in the WebGL treatment. It must be emphasized that these effect sizes are based on comparisons between very small gains. What is worth noting is that females are not disadvantaged by the VR treatment compared to other treatments.

There was some dependence on self-reported prior experience with heavily-3D videogames. A repeated measures analysis in SPSS of students within a given treatment type using 3D videogame play as a between-subjects factor showed a significant difference between gains by gamers and non-gamers, as outlined in Table IV. Students who reported heavy gameplay outperformed students with low videogame play in all three treatments. But unlike the work by Smith et al. [15], the VR treatment did not benefit frequent video game players more than the WebGL and Images treatment benefited them.

TABLE IV. Gains by treatment and videogaming

| Treatment | 3D Video-game Exp. | Mean Gains | Gains 95% CI | p (high vs low) |
|---|---|---|---|---|
| VR | High | 3.0 | (-0.6, 6.6) | <0.01 |
|  | Low | -0.4 | (-4.1, 3.2) |  |
| WebGL | High | 4.1 | (0.6, 7.5) | <0.01 |
|  | Low | -2.1 | (-5.9, 1.6) |  |
| Images | High | 4.7 | (0.9, 8.4) | <0.01 |
|  | Low | 1.9 | (-1.9, 5.8) |  |

### C. Factors that may contribute to poor gains

The low gains in the Images treatment are to be expected, since students were only shown images very similar to those they had already seen multiple times in their classes, homework, and textbooks. The fact that no additional gains were observed in the VR and video treatments may be explained by the fact that these treatments were not highly interactive. The potential importance of this is discussed further in the Conclusions section.

The fact that the assessment items were experimental introduces two potential problems: (1) Many items did not load well onto subscales as theoretically predicted. (2) The assessments were long, and test fatigue may have contributed to flat performances.

### IV. CONCLUSIONS

We have conducted a controlled study of VR treatments in teaching freshman magnetostatics. The overall low gains and relative independence of gains on treatment type are similar to those found in other VR studies. Work is currently under way to make VR interventions that are much more constructively interactive through the use of a Bluetooth controller paired to the phone, allowing more diverse input. However, it is worth noting that recently Madden et al. [14] found a similar lack of dependence of gains on treatment when using an augmented reality (AR) treatment, which they argue is more interactive than the treatment used in Smith et al., and likely more than the present work. They found this to be the case, even when students showed a strong attitudinal preference for AR over other media. Additionally, Brown et al. found very small differences between VR and control treatments in a highly interactive intervention in an introductory engineering context [22]. More work is needed to establish the importance of constructive interactivity in VR/AR interventions.

The preliminary assessment used in this study is a useful initial step toward developing a heavily 3D conceptual magnetostatics instrument. Significant additional work is needed. Future work may build upon the subscale assessing the magnetic fields due to two current sources, the only subscale with an acceptably high Cronbach's alpha.

In these data there is no indication that VR treatments disadvantage one sex over another. Students reporting heavy 3D videogame-play do experience higher gains. This holds for all treatment groups, unlike the previous study by Smith et al. [15] in which only VR-treated gamers had significantly higher gains than non-gamers. If gaming is a reasonable proxy for familiarity with visuospatial rotations in an electronic context, it is not clear why it previously correlated with higher gains in the VR treatment only, and now correlates with higher gains in all treatments. Future work will not rely solely on videogame play as an indicator, but will also examine data from students given a preliminary training in the VR environment, prior to instruction. Future studies might cross-reference prior videogame play with scores on validated metrics such as the Purdue Visualization of Rotations Test [23].

### ACKNOWLEDGMENTS

The development of VR visualizations has been supported by the OSU STEAM Factory, OSU's Marion campus, and the Office for Distance Education and eLearning (ODEE) at OSU. The WebGL websites were hosted by OSU's ASCTech.